\documentclass[sigplan,nonacm]{acmart}
\usepackage[linesnumbered,ruled,vlined,noend]{algorithm2e}
\DontPrintSemicolon

\SetKwComment{Comment}{\color{green!50!black}// }{}

\newcommand{\assign}{\leftarrow}
\newcommand{\var}{\texttt}
\newcommand{\typehint}{\textbf}
\newcommand{\FuncCall}[2]{\texttt{\bfseries #1(#2)}}
\SetKwProg{Function}{function}{}{}
\SetKw{kwMax}{max}
\SetKw{KwBreak}{break}

\usepackage{pifont}

\usepackage{tikz}
\usepackage{xcolor}
\usepackage{amsmath}
\usepackage{mathtools}
\usepackage{graphicx}
\usepackage{svg}
\usepackage{colortbl}
\usepackage{tabu}
\usepackage{listings}
\usepackage{verbatim}
\usepackage{lipsum}
\usepackage{todo}

\usepackage{cleveref}

\usepackage{titlesec}
\usepackage{booktabs}
\usepackage{graphics}
\usepackage{arydshln}
\PassOptionsToPackage{table}{xcolor}
\usepackage{graphicx}
\usepackage{makecell}
\usepackage{enumitem}
\setlist[enumerate,1]{nosep, leftmargin=*}
\setlist[itemize,1]{nosep, leftmargin=*}

\usepackage{tikz}
\usetikzlibrary{positioning}
\usetikzlibrary{fit}
\usepackage{afterpage}

\usepackage{subcaption}
\usepackage{caption}

\usepackage{stfloats}

\usetikzlibrary{trees}

\tikzstyle{level 1}=[level distance=3cm, sibling distance=2cm]
\tikzstyle{level 2}=[level distance=3cm, sibling distance=1cm]

\tikzstyle{bag} = [text width=1em, text centered]
\tikzstyle{end} = [circle, minimum width=3pt,fill, inner sep=0pt]

\graphicspath{ {./src/} }

\definecolor{Gray}{gray}{0.9}

\definecolor{theme-green}{HTML}{89b374}
\definecolor{theme-darkblue}{HTML}{384890}
\definecolor{theme-grey}{HTML}{e4e4e4}
\definecolor{theme-darkgrey}{HTML}{808080}
\definecolor{theme-blue}{HTML}{387490}
\definecolor{theme-purple}{HTML}{903874}

\usepackage{accsupp}
\newcommand{\noncopynumber}[1]{%
	\BeginAccSupp{method=escape,ActualText={}}%
	\tiny\color{theme-darkgrey}
    #1%
	\EndAccSupp{}%
}

\DeclareCaptionFormat{listing}{}

\author{Hugo Pompougnac}
\affiliation{\institution{Univ. Grenoble Alpes, Inria, CNRS, Grenoble INP, LIG, 38000 Grenoble, France} \country{}}
\email{hugo.pompougnac@inria.fr}

\author{Alban Dutilleul}
\affiliation{\institution{ENS Rennes} \country{France}}
\email{alban.dutilleul@ens-rennes.fr}

\author{Christophe Guillon}
\affiliation{\institution{Univ. Grenoble Alpes, Inria, CNRS, Grenoble INP, LIG, 38000 Grenoble, France} \country{}}
\email{christophe.guillon@inria.fr}

\author{Nicolas Derumigny}
\affiliation{Télécom SudParis \country{France}}
\email{derumigny.nicolas@telecom-sudparis.eu}

\author{Fabrice Rastello}
\affiliation{\institution{Univ. Grenoble Alpes, Inria, CNRS, Grenoble INP, LIG, 38000 Grenoble, France} \country{}}
\email{fabrice.rastello@inria.fr}

\title{Performance bottlenecks detection through microarchitectural sensitivity}

\begin{document}

\date{}

\newcommand{\toolsensitivity}{\texttt{Gus}}
\newcommand{\qemu}{\texttt{QEMU}}
\newcommand{\tcg}{\texttt{TCG}}
\newcommand{\ecm}{\texttt{ECM}}
\newcommand{\uopsinfo}{\texttt{UOPS.INFO}}
\newcommand{\callgrind}{\texttt{callgrind}}
\newcommand{\cachegrind}{\texttt{cachegrind}}
\newcommand{\tstart}{\texttt{t\textsubscript{start}}}
\newcommand{\tend}{\texttt{t\textsubscript{end}}}
\newcommand{\tmin}{\texttt{t\textsubscript{min}}}
\newcommand{\tmax}{\texttt{t\textsubscript{max}}}
\newcommand{\tavail}{\texttt{t\textsubscript{avail}}}
\newcommand{\instrlat}{\texttt{instruction\textsubscript{latency}}}
\newcommand{\valgrind}{\texttt{valgrind}}
\newcommand{\llvmmca}{\texttt{llvm-mca}}
\newcommand{\toplev}{\texttt{TopLev}}
\newcommand{\iaca}{\texttt{iaca}}
\newcommand{\vtune}{\texttt{VTune}}
\newcommand{\perf}{\texttt{perf}}
\newcommand{\perfcmd}{\texttt{perf stat}}
\newcommand{\facile}{\texttt{facile}}
\newcommand{\uica}{\texttt{uica}}
\newcommand{\osaca}{\texttt{osaca}}
\newcommand{\granite}{\texttt{granite}}
\newcommand{\ithemal}{\texttt{ithemal}}
\newcommand{\difftune}{\texttt{difftune}}
\newcommand{\cesasme}{\texttt{CesASMe}}

\newcommand{\ie}{\textit{i.e.}}
\newcommand{\etc}{\textit{etc.}}
\newcommand{\eg}{\textit{e.g.}}
\newcommand{\muop}[1]{$\mu$op#1}
\newcommand{\micro}[1]{$\mu$#1}
\newcommand{\instr}[1]{\texttt{#1}}
\newcommand{\ii}[1]{\instr{#1}}
\newcommand{\muinstr}[1]{\texttt{$\mu$-#1}}
\newcommand{\mi}[1]{\muinstr{#1}}
\newcommand{\figref}[1]{Fig.~\ref{fig:#1}}
\newcommand{\secref}[1]{Section~\ref{sec:#1}}
\newcommand{\textgray}[1]{\textcolor{gray}{#1}}
\newcommand{\hwc}[1]{\texttt{#1}}

\newcommand{\decan}{\texttt{DECAN}}
\newcommand{\maqao}{\texttt{MAQAO}}
\newcommand{\saake}{\texttt{SAAKE}}







\begin{abstract}

  Modern Out-of-Order (\textit{OoO}) CPUs are complex systems with many components interleaved in non-trivial ways.
  Pinpointing performance bottlenecks and understanding the underlying causes of program performance issues are critical tasks to make the most of hardware resources. 

  We provide an in-depth overview of performance bottlenecks in recent 
  \textit{OoO} microarchitectures and describe the difficulties of detecting them.
Techniques that \textit{measure} resources utilization can offer a good understanding of a program's execution, but, due to the constraints inherent to Performance Monitoring Units (PMU) of CPUs, do not provide the relevant metrics for each use case.

Another approach is to rely on a performance model to \textit{simulate} the CPU behavior.
Such a model makes it possible to implement any new microarchitecture-related metric.
Within this framework,
we advocate for implementing modeled resources as parameters that can be varied at will to reveal performance bottlenecks. This allows a generalization of bottleneck analysis that we call \textit{sensitivity analysis}.

We present \toolsensitivity{}, a novel performance analysis tool that 
combines the advantages of sensitivity analysis and dynamic binary instrumentation 
within a resource-centric CPU model.
We evaluate the impact of sensitivity on bottleneck analysis over a set of high-performance computing kernels.

\end{abstract}

\maketitle

\section{Introduction}\label{sec:intro}
\label{sec:intro}

The need for compute keeps growing with massive deployment of intensive
computations, whereas machines computing power remains limited by the
laws of physics~\cite{darksillicon2011}.
Thus, the optimization of compute kernels is an increasingly critical
activity. Indeed, when it comes to real-time embedded systems or
high-performance computing, a program's non-functional requirements (execution
time, memory footprint, \etc{}) are at least as important as its functional
requirements. Consequently, unexpected sources of inefficiency can be seen as
performance \textit{bugs}, which need to be corrected if the program
implementation is to meet its specification.

One way of guiding the optimization process is to inspect the Performance
Monitoring Units (PMU) of the CPU.
These hardware counters provide many measures for evaluating program
performance, including the cost of execution in CPU cycles or CPU slots.
Those can then be refined. For instance, the expert who optimizes a
basic block (a sequence of binary or assembly instructions
whose only entry point is its first instruction, and whose only exit point
is its last instruction\footnote{
Strictly speaking, a basic block includes a final branch
instruction, but during analysis, in the case of a loop body,
it is common practice to ignore the latter\cite{uica}.})
reports the \textit{throughput} of this basic block, \ie{} its cost per iteration.
Hardware counters also provide raw knowledge on
the behavior of the microarchitecture's resources.
Resources that have a limiting impact on a program performance
are known as \textit{bottlenecks} of this program and should be the focus of
optimization efforts.

\begin{figure}
  \begin{lstlisting}[language={[x86masm]Assembler},numbers=left]
mov    -0x10(%rsp),%rdx
vmovsd (%rdx,%rax,1),%xmm0
vaddsd 0x8(%rdx,%rax,1),%xmm0,%xmm0
vaddsd 0x10(%rdx,%rax,1),%xmm0,%xmm0
vmulsd %xmm1,%xmm0,%xmm0
mov    -0x18(%rsp),%rdx
vmovsd %xmm0,0x8(%rdx,%rax,1)
mov    -0x18(%rsp),%rdx
vmovsd -0x8(%rdx,%rax,1),%xmm0
vaddsd (%rdx,%rax,1),%xmm0,%xmm0
vaddsd 0x8(%rdx,%rax,1),%xmm0,%xmm0
vmulsd %xmm1,%xmm0,%xmm0
mov    -0x10(%rsp),%rdx
vmovsd %xmm0,(%rdx,%rax,1)
\end{lstlisting}
  \caption{A computational kernel implementing the Jacobi iteration (vectorized)
    \label{fig:jacobi_bb}.}
\end{figure}

In a modern Out-of-Order (OoO) CPU, the components that are often subject of
bottleneckness analyses are:
\begin{itemize}
\item The \textbf{front-end}, which embeds the instruction decoding process and
  produces sequences of micro-operations (\muop{s}) from the input program.
\item The \textbf{back-end}, or execution engine,
  which embeds the core's computing units as well as 
  the execution ports through which \muop{s}
  access the formers; the Allocator which decides which \muop{}
  should be sent to which port; and the Scheduler which retains \muop{s}
  (coming from the front-end) until
  their dependencies have been resolved.
\item The \textbf{speculative execution}, a mechanism based on the prediction of
 branch conditions that executes instructions belonging to uncertain code paths. In case of mispredictions, performance are degraded.
\item The \textbf{memory subsystem}, corresponding to the cache hierarchy
and the off-chip memory.
\end{itemize}

Consider the assembly basic block in \figref{jacobi_bb},
which implements the innermost loop body of a Jacobi iteration.

The instructions on lines 7 and 14 are the only ones
that do not load the Address Generation Units (AGU), while the others
systematically manipulate the pointers stored in \texttt{\%rsp} and \texttt{\%rdx}. On
Intel Skylake microarchitecture
(produced until 2022 under the \emph{Comet Lake} family),
there are only two general purpose AGUs: behind ports 2 and 3.
Consequently, repeated execution of this basic block is likely to saturate
them. Assuming that the calculation is L1-resident, and given that
the surrounding \texttt{for} loops (not shown here) have predictable control flow,
these ports are the most likely the bottleneck, which translates to a back-end cause.

The state-of-the-art Topdown Analysis Method (TAM)\cite{topdown}
exploits the Performance Monitoring Units (PMUs) through hardware counters
to identify bottlenecks, and is the approach used in Intel's own performance
debugging tool, \vtune{}~\cite{vtune}. Detailed values of individual counters
is also available for analysis through Linux \perf{}\cite{perf}.
In TAM, pipeline slots describe the number of \muop{s} a microarchitecture is capable of processing during a time interval. A slot is \emph{filled} each time a \muop{} has been successfully processed (\emph{retired}). 
A bottleneck is then determined as resource that causes \emph{a sufficient number}
of slots to be unused; thresholds being set heuristically.

This technique is illustrated in \figref{jacobi_perf} which
reproduces a subset of the \perfcmd{} report over
the basic block from \figref{jacobi_bb} in realistic
conditions (being iterated thousands of times).
Here, every slot is classified as unused because of
the speculative execution (line 11), the front-end (line 12), the back-end
(line 13), or having been filled (line 10): as expected, the back-end is detected as the bottleneck.


In most cases, TAM provides sufficient knowledge to guide optimization. 
However, fine-grain PMU-based analysis is known to be limited~\cite{weaver2013non},
as hardware counters traditionally relies on interrupt or event-based sampling.
In practice, this means that the association from saturated resources to saturating
assembly instructions is often lost, leading to tedious performance debugging.

This led to the development of a complementary approach: binary code analysis,
based on the computation of metrics through a performance model at assembly level.
Although these approaches produce estimations necessarily less precise than measurements,
performance models offer a wider range of metrics and greater flexibility in
analysis than hardware counters, and do not require any hardware support.
%

\begin{figure}
\begin{subfigure}{\linewidth}
  \footnotesize
  \centering
    \begin{tabular}{ l l l l }
      \textgray{10}  & 191281317 & td-retiring & 39,6\% Retiring \\
      \textgray{11}  & 11363246 & td-bad-spec & 2,4\% Bad Speculation \\
      \textgray{12}  & 17044869 & td-fe-bound & 3,5\% Frontend Bound \\
      \textgray{13}  & 263248545 & td-be-bound & 54,5\% Backend Bound \\
    \end{tabular}
  \caption{}
  \label{fig:jacobi_perf}
\end{subfigure}

\begin{subfigure}{\linewidth}
	\centering
  	\footnotesize
    \begin{tabular}{ l l | c | c | c | c | c | c  }
      \multicolumn{2}{l}{Instructions} & \muop{s} & p0 & p1 & p2 & p3 & p4  \\
      \hline
      \textgray{1} & \instr{mov}     & 1 &   &   & 1 &   &   \\
      \textgray{2} & \instr{vmovsd}  & 1 &   &   &   & 1 &   \\
      \textgray{3} & \instr{vaddsd}  & 2 & 1 &   & 1 &   &   \\
      \textgray{4} & \instr{vaddsd}  & 2 & 1 &   & 1 &   &   \\
      \textgray{5} & \instr{vmulsd}  & 1 & 1 &   &   &   &   \\
      \textgray{6} & \instr{mov}     & 1 &   &   & 1 &   &   \\
      \textgray{7} & \instr{vmovsd}  & 2 &   &   &   & 1 & 1 \\
      \textgray{8} & \instr{mov}     & 1 &   &   & 1 &   &   \\
      \textgray{9} & \instr{vmovsd}  & 1 &   &   &   & 1 &   \\
      \textgray{10} & \instr{vaddsd} & 2 &   & 1 & 1 &   &   \\
      \textgray{11} & \instr{vaddsd} & 2 & 1 &   &   & 1 &   \\
      \textgray{12} & \instr{vmulsd} & 1 &   & 1 &   &   &   \\
      \textgray{13} & \instr{mov}    & 1 &   &   & 1 &   &   \\
      \textgray{14} & \instr{vmovsd} & 2 &   &   &   & 1 & 1 \\
    \end{tabular}
  \caption{}
  \label{fig:jacobi_iaca}
\end{subfigure}

\label{fig:jacobi_iaca_perf}
\caption{Extract of \perfcmd{} (\subref{fig:jacobi_perf}) and \iaca{} (\subref{fig:jacobi_iaca}) outputs when run on the basic block from \figref{jacobi_bb}.}


\end{figure}

Intel \iaca{} was the code analyzer distributed by Intel until 2019.
When applied to the basic block in \figref{jacobi_bb}, it delivers a report
including a throughput estimation, a detailed ports pressure analysis
and a bottleneck analysis which outputs the most probable bottleneck cause.
\figref{jacobi_iaca} shows a simplified transcription of the instruction-centered ports 
pressure table for \figref{jacobi_bb}. 
It confirms our initial intuition that most instructions use
ports 2 and 3 of a Skylake architecture.


In \iaca{}, bottleneck analysis essentially consists in looking for the source of
probable \textit{stalls}. Stalling occurs when a pipeline slot is empty:
the corresponding hardware resources are not fed, and therefore
not used during one cycle.
In this way, \iaca{} seeks either to identify a saturated component (front-end or back-end),
or to detect dependencies between \muop{s} which would result in stalls in the Scheduler (latency bound). 
Other code analyzers (\uica{}\cite{uica}, \llvmmca{}\cite{llvmmca}\ldots{}) rely on different heuristics,
but all embed such performance models.

We propose to generalize this approach with a method called \emph{sensitivity}, that consists in running several
simulations of the same program with different \emph{accelerations} of CPU resources. In this framework, 
accelerating a resource means varying its ability to process \muop{s} (e.g. increasing the throughput of one port).
This allows us to design a new metric: the global speed-up caused by the acceleration of a resource.
When the speed-up is strictly positive, the resource is a bottleneck and should focus the optimization efforts.

Simulated resources can be considered at any level of granularity. 
Let us go back to the example in \figref{jacobi_bb}. 
Similarly to TAM and \iaca{}, a sensitivity-based approach concludes that the back-end is bottleneck. 
Without any further information, a more precise report requires an expert to analyze the assembly code. 
But, by simply accelerating ports 2 and/or 3, sensitivity analysis can immediately conclude that the bottleneck is at the
address calculation level.
In this paper, we present a tool called \toolsensitivity{} implementing such a sensitivity-oriented analysis.


Our contributions are presented as follows:
\begin{itemize}
\item In \secref{bottleneckness}, we describe the challenges linked with bottleneck identification; we propose a new definition of a bottleneck based on sensitivity analysis and show how this technique generalize existing approaches.
\item In \secref{gus}, we present the code analyzer \toolsensitivity{}. We detail its underlying performance model and how it implements sensitivity analysis.
\item In \secref{experiments}, we evaluate our approach on a set of high-performance computing kernels.
\end{itemize}

\section{Understanding bottlenecks}\label{sec:bottleneckness}

Identifying the hardware resource (resp. resources) that is (resp. 
are) bottleneck for running a given program on a given microarchitecture is
a difficult problem.  Indeed, the construction of such information implies
the aggregation and weighting of estimates relating to heterogeneous and
intertwined components that are already difficult to understand separately,
and even more so together.  In order to build a performance model capable of
highlighting such bottlenecks, it is therefore important to first acquire a
good understanding of the microarchitecture, even at high level.


\subsection{An Out-of-Order CPU architecture}

\begin{figure}
  \begin{center}
    \includesvg[inkscapelatex=false, width=\linewidth]{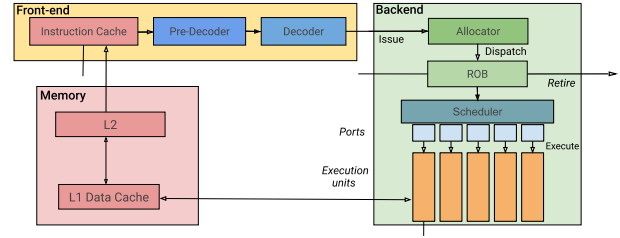}
  \end{center}
  \caption{Simplified view of a pipelined OoO CPU core.}
  \label{fig:cpu}
\end{figure}

\figref{cpu} is a high-level simplified representation of a pipelined modern
out-of-order CPU core.
During its lifecycle, each \muop{} passes through 4 distinct states: it is
\textit{issued} by the front-end, \textit{dispatched} by the allocator,
\textit{executed} by the execution units during its stay in the ROB,
and eventually \textit{retired} once its effects have been committed.

\paragraph{Front-end}

The Branch Prediction Unit (BPU) guides instruction fetching with a given
Program Counter (PC), potentially speculating on the outcome of a branch. 
The instruction fetching unit then typically reads a binary chunk from the
Instruction Cache (IC). 
The Pre-Decoder is thereafter responsible for translating this binary chunk
into instruction(s), which are themselves broken down to micro-operations
(\muop{s}) by the decoding unit. The decoded \muop{s} are filled into
the Instruction Decode Queue (IDQ, also called Allocation Queue)
while awaiting their delivery to the back-end.

\paragraph{Allocator/Renamer}

Each \muop{} produced by the front-end is first filled into a pipeline slot and assigned to an execution port by the Allocator\cite{intel}. 
Since different ports can have the same role, a given \muop{} can be assigned to any of these ports.
This allocation mechanism is illustrated in \figref{allocation}, where a basic block consisting of the instruction sequence \instr{mul} \instr{sbb} \instr{rol} \instr{bsf} \instr{rol} \instr{pop} \instr{mov} \instr{sahf} \instr{movsx} \instr{sahf} \instr{sbb} \instr{xor} is scheduled on an Intel Skylake CPU (8 ports). 
For the sake of clarity, all of these instructions produces one \muop{}, and are assumed not to have dependencies.
At the same time as allocation, registers are renamed, ensuring mapping
between logical and physical registers. This operation is controlled by the
Register Alias Table (RAT, also called Resource Allocation Table).

\begin{figure}
  \footnotesize
  \begin{center}
    \begin{tabular}{ | l l | l | l | }
      \hline
      \multicolumn{2}{|l|}{Instructions}      & Possible ports    & Dispatched port \\
      \hline
      \textgray{1}  & \ii{mul}     & p1                               & p1 \\
      \textgray{2}  & \ii{sbb}     & p0 $\lor$ p0                     & p0 \\
      \textgray{3}  & \ii{rol}     & p0 $\lor$ p6                     & p6 \\
      \textgray{4}  & \ii{bsf}     & p1                               & p1 \\
      \textgray{5}  & \ii{rol}     & p0 $\lor$ p6                     & p6 \\
      \textgray{6}  & \ii{pop}     & p2 $\lor$ p3                     & p2 \\
      \textgray{7}  & \ii{mov}     & p2 $\lor$ p3                     & p3 \\
      \textgray{8}  & \ii{sahf}    & p0 $\lor$ p6                     & p6 \\
      \textgray{9}  & \ii{movsx}   & p2 $\lor$ p3                     & p2 \\
      \textgray{10} & \ii{sahf}    & p0 $\lor$ p6                     & p0 \\
      \textgray{11} & \ii{sbb}     & p0 $\lor$ p6                     & p6 \\
      \textgray{12} & \ii{xor}     & p0 $\lor$ p1 $\lor$ p5 $\lor$ p6 & p5 \\
      \hline 
    \end{tabular}
  \end{center}
  \caption{The allocation mechanism applied to a basic block of
    dependencies-free instructions (each being broken into
    exactly 1 \muop{}).}
  \label{fig:allocation}
\end{figure}

\paragraph{Re-Order Buffer} \label{par:rob}

The allocated \muop{s} are then used to fill the re-order buffer (ROB), which
exploits instruction-level parallelism by enabling OoO execution. 
The ROB is a circular buffer that is filled incrementally, with \muop{s} being
issued at the \emph{tail} and retired from the \emph{head}.
Retiring a \muop{} from the ROB means that its effects on the architectural
state of the pipeline are committed.
The execution can be done out-of-order to increase the utilization of the
execution units; two \muop{s} can be executed in parallel if they rely on two
different and available resources and if they are not dependant on each other. 
However, retiring \muop{s} happens in-order to preserve the illusion of
sequential execution. This idea is an extension of Tomasulo's
algorithm~\cite{tomasulo1967}.

\begin{figure}
  \footnotesize
  \begin{center}
    \begin{tabular}{ | l | l | l | l | }
      \hline
         Time & ROB             & \muop{} dispatch & \muop{} state \\
     \hline
            1 & \muinstr{mul}   & p1                & D $\rightarrow$ R \\
              & \muinstr{sbb}   & p0                & D $\rightarrow$ R \\
              & \muinstr{rol}   & p6                & D $\rightarrow$ R \\
              & \muinstr{bsf}   & p1                & D (no change) \\
      \hline
            2 & \muinstr{bsf}   & p1                & D $\rightarrow$ R \\ 
              & \muinstr{rol}   & p6                & D $\rightarrow$ R \\
              & \muinstr{pop}   & p2                & D $\rightarrow$ R \\
              & \muinstr{mov}   & p3                & D $\rightarrow$ R \\
      \hline
            3 & \muinstr{sahf}  & p6                & D $\rightarrow$ R \\ 
              & \muinstr{movsx} & p2                & D $\rightarrow$ R \\ 
              & \muinstr{sahf}  & p0                & D $\rightarrow$ R \\ 
              & \muinstr{sbb}   & p6                & D (no change) \\
      \hline
            4 & \muinstr{sbb}   & p6                & D $\rightarrow$ R \\
              & \muinstr{xor}   & p5                & D $\rightarrow$ R \\
              & -               & -                 & - \\ 
              & -               & -                 & - \\ 
            \hline
    \end{tabular}
    \end{center}
    \caption{
      A ROB of size 4 processing the basic block
      in \figref{allocation} along time, where
      \texttt{$\mu$-instr} designates the
      \muop{} 
      decoded from the instruction \texttt{instr}.
      "\muop{} state" column describes changes between D(ispatched) and
      R(etired) \muop{} state.
        \label{fig:rob}}
\end{figure}

\figref{rob} illustrates this mechanism with a simplified ROB of 4 entries
(instead of 224 in a Skylake micro-architecture), allowing 4 \muop{s} to be issued and retired per cycle. 
The latter is fed by the decoding of the basic block shown in \figref{rob}:
\begin{itemize}
\item At cycle 1, the first 4 \muop{s} are dispatched.
  The \muop{s} \muinstr{mul}, \muinstr{sbb} and \muinstr{rol}
  are simultaneously executed by ports 1, 0 and 6.
  The \muop{} \muinstr{bsf} being mapped on the same port than
  \muinstr{mul} (\ie{} the port 1), it can not be executed for now.
  The 3 executed \muop{s} are then retired.
\item At cycle 2, the following 3 \muop{s}
  are dispatched. All the \muop{s} contained in the ROB being mapped on different
  ports (\ie{} the ports 1, 6, 2 and 3), they are executed in parallel
  and then retired.
\item At cycle 3, the following 4 \muop{s} are dispatched.
  The \muop{s} \muinstr{sahf}, \muinstr{movsx} and \muinstr{sahf}
  are simultaneously executed by ports 6, 2 and 0.
  The \muop{} \muinstr{sbb} being mapped on the same port as
  \muinstr{sahf} (\ie{} the port 6), it can not be executed for now.
  The 3 executed \muop{s} are then retired.
\item At cycle 4, the last \muop{} of the sequence, \muinstr{xor}, is
  dispatched. \muinstr{sbb} and \muinstr{xor} being mapped
  on different ports, they are executed in parallel and finally retired.
\end{itemize}

\paragraph{Scheduler/Reservation Station}

Executing a \muop{} means transmitting it to the Scheduler, also known as
Reservation Station.  The latter holds each \muop{} until the execution port
to which it is assigned is free and its dependencies are satisfied. 
Then, it is executed by its mapped port.

\paragraph{Execution ports and execution units}
Each port is linked to execution units responsible for a class of operations: 
arithmetic calculation and logic operations (ALU), divider, memory reads,
memory writes, vector calculation, \etc{} A port processes a maximum of
one \muop{} per cycle and is said to be saturated if it does not find itself
waiting and can actually consume a new \muop{} as soon as it has finished with
the previous one.

\paragraph{Memory communications}

During their execution, load-related \muop{s} are stored in Load Buffers
(and store-related \muop{s} in Store Buffers). 
The execution units processing these memory-related \muop{s}
(\eg the Load-Store Units) have to wait for the data on which they operate,
which induces \textit{latencies}. 
Depending on the location of the required data, it must be incrementally
repatriated from the farthest memory to the nearest one.

Repatriated or newly-computed data is stored in processor registers, can be vector or integer. 
Each kind comes with a dedicated file included within the Scheduler and which indicates their occupancy level; 
for example, the Vector Register File indicates which vector registers are available and which are occupied.

In addition to registers, L1 and L2 caches are core-specific, while L3
cache and RAM are system-wide unified memories.

\subsection{Bottleneck analysis in state-of-the-art}

Some code analyzers do not provide bottleneck analysis and are limited to throughput estimation. 
It is generally the case of neural-network based ones such as \granite{}\cite{granite} and \ithemal{}\cite{ithemal}. 
Others, without being black boxes, only partially model the microarchitecture and are therefore specialized
for particular classes of programs.
For instance, \osaca{}\cite{osaca} assumes that a basic block is never front-end bound.
It produces, in addition to a port saturation analysis, 
a dependencies-based critical path analysis.

However, even less limited approaches, based on TAM or on code analyzers
like \iaca{} or \uica{}, lack generality. They provide bottleneck analysis
consisting in identifying under- or overexploited resources,
possibly extended to include time lost due to bad speculation.


\setlength{\jot}{8pt}
\begin{figure}[t]
  \footnotesize
  \begin{center}
    \begin{align*}
      Retiring =\ & \frac{SlotsRetired}{TotalSlots} \\
      Front\ End\ Bound =\ & \frac{FetchBubbles}{TotalSlots} \\
      Bad\ Spec =\ & \frac{SlotsIssued - SlotsRetired + RecoveryBubbles}{TotalSlots} \\
      Back\ End\ Bound =\ & 1 - (Front\ End\ Bound\ +\ Bad\ Spec\ +\ Retiring) \\
    \end{align*}
  \end{center}
  \vspace{-0.7cm}
  \caption{The PMU-based formulas on which the Topdown Analysis Method relies at high
    level.\label{fig:topdown_formulas}}
\end{figure}

\paragraph{Load-based analysis:} \iaca{}, \llvmmca{}.
The bottleneck analysis of \iaca{} exploits the intermediate data produced during a basic block throughput estimation. 
It first looks at the status of three simulated components belonging to the back-end: 
the Reservation Station (if full, the ALUs should be saturating), the Load Buffer (if full, the AGUs should be saturating) and the Vector Register File.
One of these components being full, the back-end is bottleneck
\footnote{We have obtained this heuristic by reverse engineering the tool, but it is undocumented by Intel.}. 
Otherwise, it looks for a significant difference between the throughput prediction obtained when taking latencies into account and the throughput prediction obtained when ignoring them. 
If such a difference is found, latencies are considered to have a significant influence on the basic block performance: 
the basic block is latency-bound. 
Finally, if the bottleneck is neither back-end nor latency, then the front-end is bottleneck.

The \llvmmca{}\cite{llvmmca} code analyzer, distributed as part of the LLVM project, is based on a similar approach. 
Like \iaca{}, its bottleneck analysis engine exploits the underlying performance model used for throughput estimation. 
It provides indicators of back-end resources saturation and influence of dependencies-induced latencies on performance. 
The main difference with \iaca{} is that the analyzer simply provides these values (in greater detail than \iaca{}), without concluding that any of the related components is bottleneck. 
The conclusion is up to the expert.

\paragraph{Throughput-based analysis:} \facile{}, \uica{}.
These two code analyzers are built around the same performance model associating a throughput to each component of the CPU (like \iaca{} and \llvmmca{}, they provide a statement of the pressure on resources at the instruction level). 
They differ simply in the algorithm used to make the basic block throughput estimation, which influences the bottleneck analysis. 
On the one hand, \uica{}\cite{uica} associates the throughput of a basic block with the timestamp at which the last instruction in the (possibly iterated many times) basic block was retired. 
In this case, a bottleneck resource is one whose throughput exceeds a certain hard-coded proportion of the total system throughput (97\% for some, 98\% for others). 
On the other hand, \facile{}\cite{facile} considers that the basic block throughput is the throughput of the slowest CPU resource. With this approach, the latter is reported as the (only) bottleneck limiting the performance.

\paragraph{Top-down analysis Method:} \perf{}, \vtune{}. 
Both \perf{} and \vtune{} use sampling techniques to inspect hardware counters. 
By default, they both collect the data required by the top-down analysis and provide a summary of the latter.
At high level, it consists of classifying all the slots of an execution into 4 categories: slots lost due to bad speculation, slots lost due to back-end, slots lost due to front-end, and finally filled slots, i.e. associated with a fully executed and retired \muop{}.

\figref{topdown_formulas} shows the algorithm described in \cite{topdown}
to classify slots. Note that:
\begin{enumerate}
\item The number of filled/retiring slots is given by a dedicated counter
  (counter \hwc{uops\_retired.retire\_slots}).
\item The number of front-end bound slots is the number of ``fetch bubbles'', \ie{}
  empty slots at the front-end output causing back-end underfeeding.
  In practice, this means tracking \muop{s} not delivered from the IDQ to the RAT
  (counter \hwc{idq\_uops\_not\_delivered.core}).
\item The number of bad speculation slots is equal to the number of
    issued \muop{s} (counter \hwc{uops\_issued.any})
    which have not been eventually retired from the ROB,
  plus ``recovery bubbles'', \ie{} empty slots produced by the Allocator because of recovery after 
  bad speculation (counter \hwc{int\_misc.recovery\_cycles},
  to be multiplied by 4 to convert cycles into slots).
\item All remaining slots are back-end bound.
\end{enumerate}

If a resource loses more than a certain percentage of slots, it is considered bottleneck.
The higher the percentage of filled slots, the more optimized the code is considered to be (assuming it is vectorized).
The analysis can be carried out at a finer grained level (and, in the case of \vtune{}, is pre-configured to \cite{vtune-cookbook}) to distinguish between branch mispredictions and machine clears, front-end bandwidth and front-end latency, different execution ports utilization, \etc{}.
Even at this level, this approach makes it possible to identify \textit{where} stalling occurs most (in which part of the microarchitecture), but not necessarily 1- \textit{when} it occurs in the sense of the Program Counter, due to the inherent inaccuracy of sampling techniques\cite{pebs2020} and 2- its ultimate cause. 
In the general case, both of these aspects require (human) additional analysis.

The ad-hoc nature of these analyses lies in the fact that they essentially rely on identifying resources that saturate enough or stalls enough.
In an OoO architecture, where the exploitation of parallelism and the interdependence of resources have an essential impact on performance, such metrics are insufficient.
They provide clues to guide the eye, but do not systematically identify the limiting resource, or resources, whose optimization will improve performance.
We propose sensitivity analysis to address this limitation.

\subsection{Why sensitivity analysis?}

\begin{figure}
  \footnotesize
  \begin{center}
    \begin{tabular}{ l | c c c c }
         & 1           & 2                & 3                 & 4 \\
      \hline
      p0 & \mi{sbb}    &                  & \mi{sahf}         &  \\
      \hline
      p1 & \mi{mul}    & \mi{bsf}         &                   &  \\
      \hline
      p2 &             & \mi{pop}         & \mi{movsx}        & \\
      \hline
      p3 &             & \mi{mov}         &                   & \\
      \hline
      p5 &             &                  &                   & \mi{xor} \\
      \hline
      p6 & \mi{rol}    & \mi{rol}         & \mi{sah}          & \mi{sbb} \\
      \hline
    \end{tabular}
  \end{center}
  \caption{Port occupancy over time during the execution of basic block from \figref{allocation} assuming each port can process at most 1 \muop{} per cycle.}
  \label{fig:timing}
\end{figure}

\figref{timing} illustrates the port pressure during the execution of the basic block in \figref{allocation}, assuming the simplified ROB of size 4 shown in \figref{rob}. 
The stalling of port~0 at cycle~2 is apparent,
but beyond the fact that the size of the ROB intrinsically limits parallelism,
understanding the cause of this behavior is not trivial.
Identifying a bottleneck without sensitivity requires a detailed analysis
of the execution and an in-depth understanding of the architecture.
With an OoO architecture indeed, performance depends on the amount of
parallelism the program manages to exploit, and a saturated resource does not
automatically limit parallelism.
Although such a metric is often a relevant clue when looking for a bottleneck,
it can be ambiguous, and even misleading:
in this example, the port 6, although completely saturated,
does not drag down performance.

\begin{figure}
  \footnotesize
  \begin{center}
    \begin{tabular}{ l l c c c c c c c c }
                    &            &      &         & Acc. p0,       & Acc.        & Acc. \\
                    &            & Port & Time    & p2, p3, p5     & p6          & p1 \\
      \hline
      \textgray{1}  & \mi{mul}   & p1   & 1       & 1              & 1           & 1 \\
      \textgray{2}  & \mi{sbb}   & p0   & 1       & 1              & 1           & 1 \\
      \textgray{3}  & \mi{rol}   & p6   & 1       & 1              & 1           & 1 \\
      \textgray{4}  & \mi{bsf}   & p1   & 2       & 2              & 2           & 1 \\
      \textgray{5}  & \mi{rol}   & p0   & 2       & 2              & 2           & 2 \\
      \textgray{6}  & \mi{pop}   & p2   & 2       & 2              & 2           & 2 \\
      \textgray{7}  & \mi{mov}   & p3   & 2       & 2              & 2           & 2 \\
      \textgray{8}  & \mi{sahf}  & p6   & 3       & 3              & 3           & 2 \\
      \textgray{9}  & \mi{movsx} & p2   & 3       & 3              & 3           & 3 \\
      \textgray{10} & \mi{sahf}  & p0   & 3       & 3              & 3           & 3 \\
      \textgray{11} & \mi{sbb}   & p6   & 4       & 4              & 3           & 3 \\
      \textgray{12} & \mi{xor}   & p5   & 4       & 4              & 4           & 3 \\
      \hline
                    &            &      & 4       & 4              & 4           & 3 \\
      \hline
    \end{tabular}
  \end{center}
  \caption{The retirement timing of the sequence of \muop{s}
    produced from the basic block in
    \figref{allocation}, assuming the same 4-entries ROB as in
    \figref{rob} and \figref{timing}. In the nominal case, all execution ports
    process one \muop{} per cycle. Accelerating a port, the latter processes
    two \muop{s} per cycle.
    }
  \label{fig:sens}
\end{figure}

On the other hand, the sensitivity approach, which consists of successively accelerating different resources of the microarchitecture through new simulations, immediately reveals those on which performance depends directly. 
This is demonstrated in \figref{sens}, where it can be seen that the execution of the \muop{} sequence costs one cycle less (3 instead of 4) when port 1 is accelerated. 
The latter consuming 2 \muop{s} per cycle, \mi{mul} (line 1) and \mi{bsf} (line 4) are executed simultaneously at the cycle 1. 
Consequently, the first \mi{sahf} (line 8) can be inserted in the ROB and executed at the cycle 2. 
This means that at cycle 3:
\begin{itemize}
\item The port 6 is no longer monopolized by the first \mi{sahf} (which ran at the
  previous cycle), allowing the second one (line 10) to run.
\item An additional entry is available in the ROB, allowing the insertion
  (and execution) of \mi{xor} (line 12).
\end{itemize}

In the end, the execution costs 3 cycles instead of 4.
The stalling cycle has finally been correlated to a saturation issue, but
in the sense of the \textit{momentary} saturation of port~1 at cycle~1.



\section{Gus, a sensitivity-oriented code analyzer}\label{sec:gus}

We introduce in the following \toolsensitivity{}, an instruction driven
simulator for performance debugging. Enabling \emph{sensitivity} analysis,
\toolsensitivity{} estimates the execution time, Instructions Per
Cycle, resources occupancy (global and instruction-wise), latency for a given
program.

First, we explain the design choices that led to the creation of an efficient  
instruction driven simulator (\ding{202}). Second, we detail the underlying
performance model of \toolsensitivity{} based on abstract resources
(\ding{203}). Finally, we demonstrate the usefulness of sensitivity and how it
overcomes the shortcomings of previous bottlenecks detection techniques
(\ding{204}, \ding{205}).

\begin{figure}
    \begin{center}
      \includegraphics[width=\linewidth]{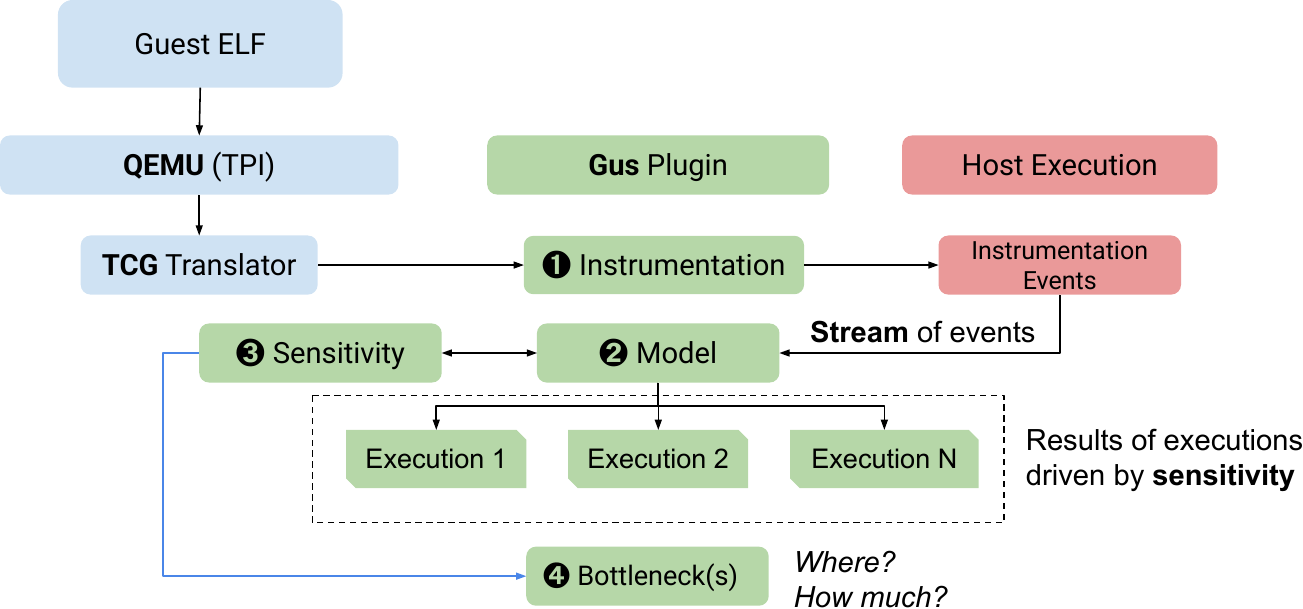}
    \end{center}
    \caption{Overview of \toolsensitivity{}'s architecture.}
    \label{fig:overview}
\end{figure}

We give a brief overview of \toolsensitivity{}'s architecture in
\figref{overview}.

\subsection{A dynamic binary instrumentation based frontend}

The front-end of \toolsensitivity{} is built around \qemu~\cite{qemu}, a fast
functional processor emulator. It relies on user-mode dynamic binary
instrumentation within a \qemu{}~plugin to generate the stream of events that
drive the simulation.

This is mostly motivated by the fact that execution traces can become
extremely large, thus storing them on disk is not usually an option for
scalability reasons.

While static code analyzers~\cite{uica,iaca} typically work on a smaller scope,
that is a single basic block, \toolsensitivity{} thanks to its trace can work on
a much larger scope chosen by the user, for instance a kernel or a whole
program.

Another advantage of having a trace is that it gives \toolsensitivity{} a lot of
information about the execution of the program, such as whether a branch was
taken or not, if two memory accesses alias, \etc{}. Static code analyzers don't
have access to such dynamic information and thus have to make assumptions that
may not hold in practice. Memory dependencies are a good example of this, as
most code analyzers ignore them completely, that leads to imprecise
results~\cite{anica}.

\paragraph{Instrumentation}
Many dynamic binary instrumentation frameworks exist, such as Pin~\cite{pin},
Valgrind~\cite{valgrind}, or DynamoRIO~\cite{dynamorio}. \toolsensitivity{} is
rather based on a modified version of \qemu{}~\cite{guillon2011}. Multiple
reasons motivated this choice, that we briefly describe here.

First, contrary to other frameworks, \qemu{} is able to cross translate between
different ISAs and application binary interfaces (ABIs). This is a major
advantage for a performance debugging tool, as it allows getting interesting
insights on micro-architectures that are not directly accessible to the user, or
even that have scarce hardware performance counters (e.g. such as a lot of ARM
or RISC-V based cores).

Second, the IR (Intermediate Representation) used by \qemu{} called \tcg{} (Tiny
Code Generator) IR is quite amenable to our needs of instrumentation as it
abstracts away the details of the underlying ISA and ABI. While Pin~\cite{pin}
and DynamoRIO~\cite{valgrind} typically introduce a much lower overhead than
\qemu{}, they do not have such an abstraction layer as the guest code is not
transformed into a very different representation.

\tcg{} is essentially a trace based JIT~\cite{mitchell1970}. Instead of
translating whole functions or the program at once, \tcg{} works on small linear
sequences of guest instructions called trace blocks. These trace blocks are
formed from the guest program using a very simple mechanism. Whenever the guest
CPU executes an instruction that has not been translated yet a new trace block
is started. This first instruction of the trace block is referred to as its
entry. \qemu{} then parses the guest instructions following the entry until it
hits a branch instruction.

The modified \qemu{} used by \toolsensitivity{} has a mechanism for loading
plugins, called the \tcg{} plugin infrastructure (TPI), that can both observe
and alter the translation process of \tcg{} IR. Such plugin can inject TCG IR
instructions into TBs before they get translated back to host machine code.

\begin{figure}
    \begin{center}
      \includegraphics[width=\linewidth]{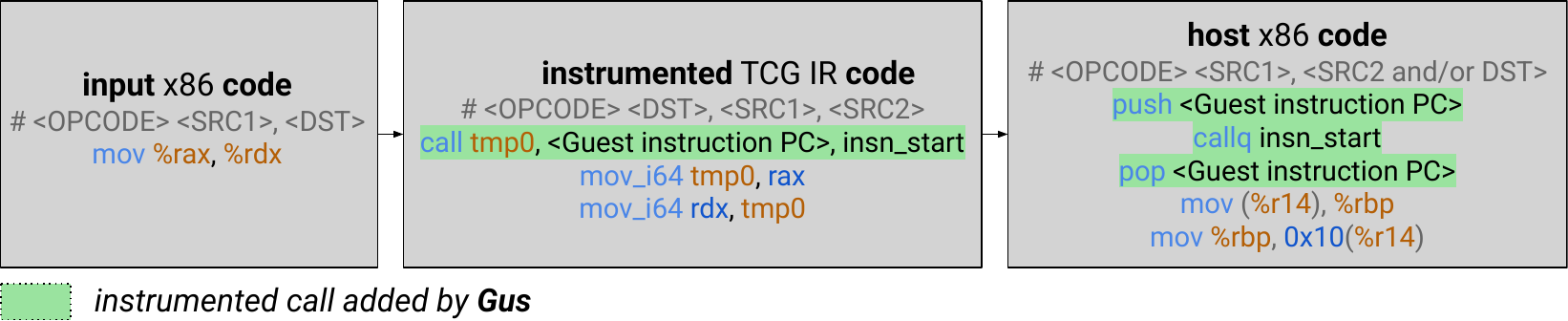}
    \end{center}
    \caption{Differents stages of a trace block in \qemu{} and the corresponding
    instrumentation generated by \toolsensitivity{}.}
    \label{fig:instrumentation}
\end{figure}

\toolsensitivity{} uses this layer to add instrumentation into the guest code.
We monitor multiple events such as when an instruction is executed, when a
memory access occurs, when a branch is taken, \etc{} as a small example in
\figref{instrumentation} demonstrates. The number of events is kept to a minimum
to reduce the overhead of the instrumentation. This instrumentation of course
requires a mapping between the guest instructions and the resulting \tcg{} IR
instructions, this is possible by disassembling the guest code with the API
provided by the TPI infrastructure.

\subsection{An abstract resource-centric performance model} 

\toolsensitivity{}'s abstract performance model describes a modern out-of-order
CPU. It consumes events fed by the instrumentation frontend and simulates the
state of this CPU. The following components are modeled:

\begin{itemize}
    \itemsep0em 
    \item A set of abstract throughput limited resources that model, amongst
    other things, the execution ports and the bandwidth between different levels
    of the cache.
    \item A finite-sized instruction window which models the \emph{ROB}.
    \item A shadow \emph{memory} and shadow \emph{register file} used to track
    data dependencies.
    \item A \emph{cache} simulator to detect cache misses.
    \item A \emph{branch prediction unit} simulator.
\end{itemize}

The underlying algorithm is described in Algorithm~\ref{alg:gus}. We will refer to
it when describing the different components of the model.

The instruction window is at the heart of the model, and dictates when an
instruction can be issued, its execution starting time \tstart{} is determined
by the resources it uses and data dependencies
(\cref{alg:tstart1,alg:tstart2,alg:tstart3}). The time it retires \tend{} is
simply defined as \(\tstart{} + \instrlat\) (\cref{alg:tend}).

\subsubsection{Abstract resources}
\label{sec:resources}

The building blocks of the model are the abstract throughput limited resources.
In order to execute an instruction, one or more of these resources may be used
and must be available. Otherwise, the instruction is stalled until all the
required resources become available. Each resource tracks the time at which it
will be available to accept another request, this is characterized by its
throughput.

The timestamp at which a resource \(R\) is available to accept an instruction is
denoted as \(R.\tavail\). Every time a resource is used its \(\tavail\) is
incremented by its inverse throughput (or \emph{gap})
(\cref{alg:cacheavail,alg:tavail}), since it determines the minimum amount of
time that must elapse between two uses of the resource.

\paragraph{Frontend}

The frontend is modeled as a single resource with a throughput of four
instructions per cycle. This is an optimistic assumption that may hide some
fine-grained bottlenecks, further work is needed to model the frontend more
precisely, as it may have been done in other works~\cite{uica}.

\paragraph{Execution ports and functional units}

The classical formalism to describe the throughput and sharing of backend
resources is a port mapping, a tripartite graph, which describes how
instructions decompose into \muop{s} and which functional units \muop{s} can
execute on.

\toolsensitivity{} instead uses a simpler two-level
representation~\cite{palmed}, called a resource mapping, where instructions are
directly associated with a list of abstract resources. To account for \muop{}
decomposition, a resource can appear in this list multiple times. This
representation avoids the need to explicitly model the port scheduling algorithm
in the backend and thus allows to achieve a lower runtime overhead.

\paragraph{Latencies}

The latencies used for execution units in our experiments are extracted from
\uopsinfo~\cite{Abel19a} and used to build a resource mapping. For bottleneck
analysis, we present latencies induced by dependencies as a separate resource.
The same approach can be found, for instance, in \iaca{}. The latter helps guide
optimization by highlighting the need to discover more parallelism, rather than
the need to relieve the load on a hardware component.

\paragraph{Instruction window} 

The out-of-order execution of instructions inside the ROB is modeled by an
instruction window, that is to say a finite-sized buffer of instructions that are
in-flight (issued but not retired).

The instruction window is bounded by the number of slots it has, thus if the
window becomes full, no other instructions can be issued until another one
retires. We denote \tmin{} as the earliest time a slot in the window will become
free (\cref{alg:tmin}).

\paragraph{Shadow memory and register file}

The shadow memory and shadow register are used to detect data dependencies
between instructions. 

For each memory cell or register they store the time at which the data in this
location will be available. The update mechanism is twofold.

First, when an instruction writes to a location, the shadow cell for that
location will be set to the time when the instruction is retired
(\cref{alg:shadowvalue}). Second, when a cache miss occurs this counts as a use
of all levels of the memory hierarchy up to the one where the miss occurred. The
location in the shadow memory corresponding to the accessed memory location is
then updated to the maximum availability of the involved resources
(\cref{alg:cacheavailread}).

\paragraph{Caches}

\toolsensitivity{} uses a fork~\cite{forkdinero} of the Dinero
IV~\cite{dineroiv} cache simulator to simulate hits and misses in the different
levels of the cache hierarchy. The PLRU (Pseudo-LRU) replacement policy from
this fork is used for all levels of the cache hierarchy, however cache
replacement policies implemented in commercial processors are a complex
topic~\cite{Abel20} and an exhaustive modelization is out of the scope of
\toolsensitivity{} that acts mostly as a generic modern out-of-order CPU model.

\paragraph{Branch prediction unit simulator}

The branch prediction unit is composed of a branch target buffer (BTB) modeled
as an associative array with an LRU replacement policy, and accompanied by an
indirect branch predictor (IBP) with a IITage scheme~\cite{Seznec2011ITTAGE} and
a conditional branch predictor (CBP) with an L-TAGE
scheme~\cite{Seznec2007LTAGE}. Findings in the
literature~\cite{rohou2015ibp,yavarzadeh:2023:half, zen2} show that such schemes
or close derivatives are actually used in \texttt{x86} commercial processors and
thus good candidates for the model of \toolsensitivity.

\subsubsection{Core performance model}

To achieve a reasonable execution time, some simplifying assumptions are made to the
model.

\begin{itemize}
    \item \toolsensitivity{} assumes that the program has regular access memory
    patterns with latencies that can be perfectly hidden by prefetching,
    similarly to what is done in the \ecm{}~\cite{hager2016, johannes2018} model.
    \item \toolsensitivity{} does not model either load/store queues or
    load-store forwarding. Thus the bandwidth between the CPU and L1
    is considered infinite and not modeled as a resource.
    \item \toolsensitivity{} also does not model execution pipeline hazards or
    operand forwarding.
\end{itemize}

This high-level model is sufficient to achieve state-of-the-art precision as
we will see in \autoref{ssec:throughput_estimation_accuracy}.

\begin{algorithm}
    \scriptsize 
    \caption{Core algorithm of \toolsensitivity{}} 
    \label{alg:gus}
    \Function{UpdateCaches(\var{i}, \var{type})}{
        \ForEach{$\var{loc} \in \FuncCall{LineAccesses}{\var{i}, \typehint{type}}$}{
            $\var{l} \assign \FuncCall{LowestCacheLevelHit}{loc}$\;

            \ForEach{$\var{l'} \in \FuncCall{CachesUpTo}{\var{l}}$}{
                $\var{shadow}[\var{loc}] \assign \kwMax{(\var{shadow}[\var{loc}], \var{l}.\tavail)}$\; \label{alg:cacheavailread}
            }

            \ForEach{$\var{l'} \in \FuncCall{CachesUpTo}{\var{l}}$}{
                \Comment{L1 Misses are considered free}
                \If{$\var{l'} = \var{l} \land \var{l'} = \typehint{L1}$}{
                    \KwBreak\;
                }
                $\var{l}.\tavail \assign \var{l}.\tavail + \var{l}.\var{gap}$\; \label{alg:cacheavail}
            }
        }
    }
    \Function{Simulate()}{
      \ForEach{$\var{i} \in T$}{
        \Comment{Only update cache loads at this point as stores may use bandwidth later}
        $\FuncCall{UpdateCaches}{\typehint{Load}}$\;
        $\tstart \assign \tmin$\; \label{alg:tstart1}
        \ForEach{$\var{loc} \in \var{i}.\var{reads}$}{
            $\tstart \assign \kwMax{(\tstart, \var{shadow}[\var{loc}])}$\; \label{alg:tstart2}
        }
        \ForEach{$\var{res} \in \var{i}.\var{resources}$}{
            $\tstart \assign \kwMax{(\tstart, \var{res}.\tavail)}$\; \label{alg:tstart3}
        }
        $\tend \assign \tstart + \var{i}.\var{latency}$\; \label{alg:tend}
        \Comment{Update execution resources}
        \ForEach{$\var{res} \in \var{i}.\var{resources}$}{
            $\var{res}.\tavail \assign \kwMax{(\tmin, \var{res}.\tavail)} + \var{res}.\var{gap}$\; \label{alg:tavail}
        }
        $\FuncCall{UpdateCaches}{\var{i},\typehint{Store}}$\;
        \ForEach{$\var{loc} \in \var{i}.\var{writes}$}{
            \Comment{Assume register renaming works perfectly}
            \If{$\var{loc} \equiv \typehint{Register}$}{
                $\var{shadow}[\var{loc}] \assign \tend$\;
            } \Else {
                $\var{shadow}[\var{loc}] \assign \kwMax{(\var{shadow}[\var{loc}], \tend)}$\; \label{alg:shadowvalue}
            }
        }
        $\var{window}.\var{push}(\tend)$\;
        $\tmin \assign \var{window}.\var{oldest}()$\; \label{alg:tmin}
        $\var{i}.\tend \assign \tend$\;
   
      }
    }
\end{algorithm}
  
\subsection{A sensitivity-oriented code analyzer} 

\paragraph{Sensitivity analysis in the broader world}
Sensitivity analysis allows \toolsensitivity{} pinpointing precisely across
instructions, the source of a performance bottleneck. We briefly describe here
how it compares with other related works and how it is implemented and used in
Gus.

Sensitivity analysis~\cite{hong_sensitivity_2018,kolia_2013}, also called
differential profiling~\cite{mckenney1995} works by executing a program multiple
times, each time varying the usage or capacity of one or more resources.
Bottlenecks can then be identified by observing by how much the change for each
resource impacts the overall performance, that is how sensitive performance is
to a given resource, and more precisely in our case a \emph{microarchitectural}
resource.

Moreover, sensitivity analysis is an automatic approach that does not require as
much expertise as other ad-hoc techniques implemented in performance debugging
tools~\cite{cqa_2014,kerncraft2017}.

While seemingly simple, this approach requires considering careful trade-offs to
make it practical. Today’s hardware does not offer many ways to easily vary the
capacity of resources, motivating the need for a model such as the one in
\toolsensitivity.

On the other hand, the precision level of the analysis is also a key factor in a
model. As a rule of thumb, the more precise the analysis, the more expensive it
is to run. \toolsensitivity{} contains an abstract high-level CPU core model that
is driven by reverse-engineered microarchitectural features \cite{palmed,
Abel19a} of modern commercial OoO cores that allows a certain degree of
generality.

Other models embedded in tools such as instruction driven
simulator~\cite{sanchez_zsim_2013} or cycle-accurate simulators~\cite{gem5}
usually feature a very detailed heavyweight CPU model. For example,
gem5~\cite{gem5} at its highest level of precision requires, on average, one
year to run a single SPEC2006 benchmark~\cite{sandberg_2015}. A common approach
to speed up simulations is to only simulate parts of a
program~\cite{wunderlich_2003} or to only simulate some aspect of the
processor~\cite{phd:nethercote,carlson2014aeohmcm} However, getting the right
level of granularity is a difficult task and requires a lot of expertise.

On the other side of the spectrum, if a model is too coarse-grained, it may lead
a user to focus on the wrong bottleneck. Moreover, if a tool does not give an
upper-bound on the optimization potential of a bottleneck, one can end up
wasting time on a bottleneck that is not worth optimizing. Hence, we believe
that these two aspects are key to the usefulness of a performance debugging
tool, as the expert-user time is a scarce resource. 

The idea of varying microarchitectural features is not new in itself either, as
it has been used by hardware designers for Microarchitecture Design Space
Exploration~\cite{ipek_asplos_2006, bai_archexplorer_2023} to guide
microarchitecture parameter tuning to explore the trade-offs amongst
performance, power, and area (PPA).

\decan~\cite{kolia_2013} is a dynamic performance analysis tool based on the
\maqao~\cite{DBCLAJ05a} binary analysis and instrumentation framework. \decan{}
finds bottlenecks by sensitivity analysis based on binary rewriting. That is,
\decan{} removes or modifies instructions in a kernel and checks which how much
each transformation affects overall performance. \decan{} measures the
performance of its modified kernels via hardware counters. The low overhead of
this approach allows it to quickly explore a large set of variants for its
sensitivity analysis. The downside of \decan{}’s approach is that its
transformations are of course not semantic preserving and can easily introduce
crashes or floating-point exceptions. Changing the semantics of a program like
this might, of course, also change its performance behavior in other subtle
ways, making it hard to verify or falsify the results produced by the tool.

The \saake{} system~\cite{hong_sensitivity_2018} is conceptually closer to
\toolsensitivity{} in its implementation of sensitivity analysis. It uses a fast
symbolic execution engine that estimates the runtime of GPU programs to drive
sensitivity analysis for finding bottlenecks. \saake{} input independent
abstract simulation works well for the simpler microarchitectures of GPUs since
they do not use out-of-order execution or speculation and handle branching
control flow using predicated execution. Since \saake{} does not actually
simulate the execution of instructions, there are several things it can not
compute that have to be provided externally.

\paragraph{Sensitivity analysis in Gus}



A resource is any finite quantity which can have an influence on the
execution time of a program. These may be quantities immediately linked
to a hardware component, such as the size of the ROB, or more abstract
resources, such as instructions latency.
We represent each of these quantities by a real number.
The sensitivity analysis implemented in \toolsensitivity{} consists in
varying the $n$ resources represented in the model during
successive execution time estimates performed on the same program $p$.
At each iteration, the capacity $c_r$ of a resource $r$ is successively
weighted (\textit{accelerated}) by real numbers $w_0,...,w_m$ to discover a $w$
that minimizes the estimation function $f_p$. In this framework, the other
resources capacities and the input program $p$ can be seen as constants.
Thus, $f_p$ is a function from real numbers to real numbers, 
associating with the value of $r$, an estimate of duration $t$.
The speed-up $s_r$ obtained by weighting $c_r$ by $w$ is therefore calculated
as follows:
\begin{equation*}
s_{w,r} = \frac{f_p(c_r)}{f_p(wc_r)} - 1
\end{equation*}

Resources whose variations result in a speedup greater than~0
(or any other \textit{ad-hoc} minimum threshold) are
bottleneck resources, and resources that reach the highest speed-ups
should be the focus of optimization efforts.

In the general case, the complexity of this algorithm is $O(n*m)$, $n$
being the number of the model resources, and $m$ the number of weight
candidates applied to each of the latter.
However, at a first glance, looking at a single weight value
may be sufficient for answering where are the bottlenecks of a program before
precisely quantifying their overall impact.

A visual representation of the results of the sensitivity analysis produced by
\toolsensitivity{} is shown below. We use a form of heat-map for the
visualization. Each bar in it represents one abstract resource. The height of
each bar and its color indicate the speedup predicted by \toolsensitivity{} if
the throughput of that resource is increased.

\begin{figure}
    \begin{center}
      \includesvg[inkscapelatex=false, width=\linewidth]{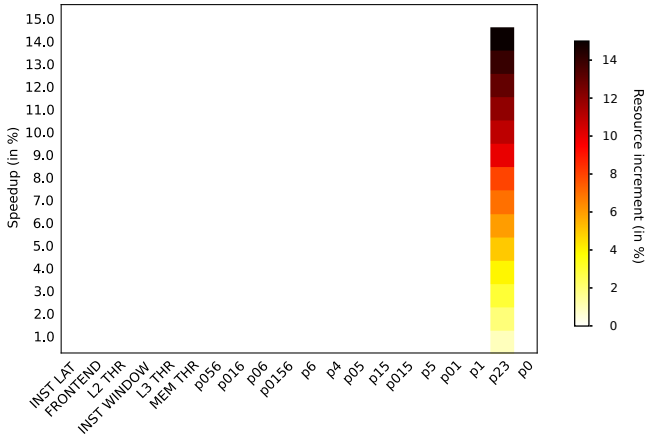}
    \end{center}
    \caption{Sensitivity analysis report produced by \toolsensitivity{}
      (Skylake architecture) for the
      Jacobi iteration method implementation discussed in
      \secref{intro}.}
    \label{fig:sensitivity}
\end{figure}

\figref{sensitivity} depicts the view of bottlenecks for the Jacobi example
 discussed in \secref{intro}.
The resource
\texttt{p23} represents the throughput of the combined execution ports 2
and 3. The graph shows that an increment of its throughput by 1\% decreases the
overall runtime of the kernel by 1\%. If we increment the resource’s throughput
by 2\%, the runtime decreases by approximately 2\%. The bar fades to black
around the 13 -- 14\% mark, which indicates that a 15\% increase in throughput
produces a 13 -- 14\% speedup. All other resources immediately fade to white at
the 0\% mark. That is, according to \toolsensitivity, the runtime of the kernel
is not affected by increasing the throughput of any other resource. Hence,
\texttt{p23} is the bottleneck of this version of the Jacobi kernel.

In addition to execution ports, resources which are subject to sensitivity
analysis in Gus are: the instruction latencies (\texttt{INST LAT}), the
front-end throughput (\texttt{FRONTEND}), the size of the ROB (\texttt{INST
WINDOW}), and the communication bandwiths between the different memory levels
(\texttt{L2 THR}, \texttt{L3 THR} and \texttt{MEM THR}).

\paragraph{Instruction-level analysis}
In a second step, one can look at the resource usage per instruction to
get a more precise idea on where is the bottleneck in the kernel.

\begin{table*}[!bt]
    \centering
    \resizebox{\textwidth}{!}{
    \begin{tabular}{lllllllllllll}
        \toprule
          PC &                                             ASM & L2 & p056 & p016 & p06 & p0156 &  p4 & p015 & p01 & p23 & F-E &                 LAT/PORTS \\
        \midrule
        12bb &                 mov rdx, qword ptr [rsp - 0x10] & 0\% &   0\% &   0\% &  0\% &    0\% &  0\% &   0\% &  0\% & \cellcolor{orange!25} 10\% &  5\% &                     2/p23 \\
        12c0 &              vmovsd xmm0, qword ptr [rdx + rax] & 0\% &   0\% &   0\% &  0\% &    0\% &  0\% &   0\% &  0\% & \cellcolor{orange!25} 10\% &  5\% &                     4/p23 \\
        12c5 &    vaddsd xmm0, xmm0, qword ptr [rdx + rax + 8] & 0\% &   0\% &   7\% &  0\% &    5\% &  0\% &   7\% & 10\% & \cellcolor{orange!25} 10\% &  5\% & 4/p016 p01 p015 p0156 p23 \\
        12cb & vaddsd xmm0, xmm0, qword ptr [rdx + rax + 0x10] & 0\% &   0\% &   7\% &  0\% &    5\% &  0\% &   7\% & 10\% & \cellcolor{orange!25} 10\% &  5\% & 4/p016 p01 p015 p0156 p23 \\
        12d1 &                         vmulsd xmm0, xmm0, xmm1 & 0\% &   0\% &   7\% &  0\% &    5\% &  0\% &   7\% & 10\% &  0\% &  5\% &     4/p016 p01 p015 p0156 \\
        12d5 &                 mov rdx, qword ptr [rsp - 0x18] & 0\% &   0\% &   0\% &  0\% &    0\% &  0\% &   0\% &  0\% & \cellcolor{orange!25} 10\% &  5\% &                     2/p23 \\
        12da &          vmovsd qword ptr [rdx + rax + 8], xmm0 & 1\% &   0\% &   0\% &  0\% &    0\% & 20\% &   0\% &  0\% &  0\% &  5\% &                      4/p4 \\
        12e0 &                 mov rdx, qword ptr [rsp - 0x18] & 0\% &   0\% &   0\% &  0\% &    0\% &  0\% &   0\% &  0\% & \cellcolor{orange!25} 10\% &  5\% &                     2/p23 \\
        12e5 &          vmovsd xmm0, qword ptr [rdx + rax - 8] & 0\% &   0\% &   0\% &  0\% &    0\% &  0\% &   0\% &  0\% & \cellcolor{orange!25} 10\% &  5\% &                     4/p23 \\
        12eb &        vaddsd xmm0, xmm0, qword ptr [rdx + rax] & 0\% &   0\% &   7\% &  0\% &    5\% &  0\% &   7\% & 10\% & \cellcolor{orange!25} 10\% &  5\% & 4/p016 p01 p015 p0156 p23 \\
        12f0 &    vaddsd xmm0, xmm0, qword ptr [rdx + rax + 8] & 0\% &   0\% &   7\% &  0\% &    5\% &  0\% &   7\% & 10\% & \cellcolor{orange!25} 10\% &  5\% & 4/p016 p01 p015 p0156 p23 \\
        12f6 &                         vmulsd xmm0, xmm0, xmm1 & 0\% &   0\% &   7\% &  0\% &    5\% &  0\% &   7\% & 10\% &  0\% &  5\% &     4/p016 p01 p015 p0156 \\
        12fa &                 mov rdx, qword ptr [rsp - 0x10] & 0\% &   0\% &   0\% &  0\% &    0\% &  0\% &   0\% &  0\% & \cellcolor{orange!25} 10\% &  5\% &                     2/p23 \\
        12ff &              vmovsd qword ptr [rdx + rax], xmm0 & 0\% &   0\% &   0\% &  0\% &    0\% & 20\% &   0\% &  0\% &  0\% &  5\% &                      4/p4 \\
        1304 &                                   add rax, 0x18 & 0\% &   0\% &   0\% &  0\% &    0\% &  0\% &   0\% &  0\% &  0\% &  5\% &                        1/ \\
        1308 &                                    cmp rax, rcx & 0\% &   0\% &   0\% &  0\% &    5\% &  0\% &   0\% &  0\% &  0\% &  5\% &                   1/p0156 \\
        130b &                                      jne 0x12bb & 0\% &   7\% &   7\% & 10\% &    5\% &  0\% &   0\% &  0\% &  0\% &  5\% &     1/p056 p016 p06 p0156 \\
        \bottomrule
        \end{tabular}
    }
    \caption{Fine grain report analysis of resource usage by instruction by \toolsensitivity{} on the \texttt{Jacobi} kernel (Skylake architecture). Resources with only zero usage across all instructions are omitted. The cells colored in orange indicate the bottleneck port.}
    \label{tab:raw-report} 
\end{table*}

This is done by looking at the fine grain report shown in \cref{tab:raw-report}.
We see that multiple instructions use the resource \texttt{p23} by loading from
memory. Reducing this, for instance by applying a register tiling
transformation, would reduce the pressure on this resource and thus improve the
overall performance of the program.

This level of precision is as we believe particularly useful for performance
debugging, as it allows a level of granularity in the general case that is not
possible with other state-of-the-art tools using for example performance
counters~\cite{topdown}.

\section{Experiments}\label{sec:experiments}

Our experiments aim first at evaluating the accuracy of the throughput
estimation tools including \toolsensitivity{} to assess the relevance of the
underlying models. We believe that having a good model is a prerequisite to
performing accurate bottleneck analysis. We then compare the bottleneck
analysis capabilities of \toolsensitivity{} to other state-of-the-art tools.

\subsection{Benchmarking framework}

Our benchmarking framework relies on \cesasme{}\cite{cesasme} and
PolyBench~\cite{polybench}.

\cesasme{} is a benchmarking harness designed to generate a large number
of microbenchmarks from an existing benchmark suite, and to evaluate
the ability of code analyzers to estimate their execution time.

Polybench is a well-known benchmark suite within the compiler
community. It contains a range of thirty numerical computations kernels with
static control flow, extracted from operations in various application domains
(linear algebra computations, image processing, physics simulation, dynamic
programming, statistics, etc.) that have attracted particular optimization
efforts.

Performance models have been shown to yield higher error rates for throughput
prediction on PolyBench than on SPEC2017~\cite{palmed}, making it a challenging
dataset for such tools.
In order to maximize diversity in the studied programs and thus
to stress different aspects of the performance models, we use
\cesasme{} to generate
1421 unique microbenchmarks from this initial suite using a three stages
process:
\begin{enumerate}
\item We use C loop nest optimizers
  --~Pluto\cite{pluto} and PoCC~\cite{pocc}~-- to generate different versions
  of each benchmark gathered in PolyBench, each coming from a different
  high-level loop optimisation~\cite{loopoptim}:
  tiling, loop fusion, unroll and jam, \etc{}
  Some transformations have no impact on the input benchmark; for example,
  loop fusion applied to a program containing a single loop. In this case,
  the fresh version of the benchmark, identical to the former one,
  is discarded.
\item We extract the innermost loop (the kernel) from each of the generated benchmark
  and discard its surrounding loop nest to prevent it
  from being memory-bound (due to
  the memory hierarchy not being taken into account by \iaca{}, \uica{}, \ldots{})
  and to limit the execution time of \perf{} and \toolsensitivity{}.
  Similarly, we use \valgrind{}~\cite{valgrind} to
  discard benchmarks whose inner loop is not L1-resident.
\item We introduce even more combinatorics by generating different binaries
  from each resulting C file using
  \texttt{gcc} options (auto-vectorization, unrolling, etc.).
\end{enumerate}

Since \toolsensitivity{} can analyze a whole function while other code analyzers
can only work on one basic block at a time, we also use the method detailed in
\cite{cesasme} to lift the
estimates they produce. In practice, we sum up the results
obtained on each basic block traversed by the program's control flow, which provides an estimate
for the whole microbenchmark.

\subsection{Throughput estimation accuracy}
\label{ssec:throughput_estimation_accuracy}

We compare the measurements on the Skylake microarchitecture provided by the
hardware counters \footnote{ On a Dell PowerEdge C6420 machine (Grid5000
cluster~\cite{grid5000}), equipped with two Intel Xeon
Gold 6130 CPUs (x86-64, Skylake microarchitecture). } and the
\toolsensitivity{} estimate with the following code analyzers:
\llvmmca~\cite{llvmmca} (\texttt{v13.0.1}), \uica~\cite{uica} (\texttt{commit 9cbbe93}), \iaca~\cite{iaca}
(v3.0-28-g1ba2cbb),
\ithemal~\cite{ithemal} (\texttt{commit b3c39a8}). 

\begin{figure}[t]
\begin{subfigure}{\linewidth}
    \centering
    \resizebox{\columnwidth}{!}{%
    \begin{tabular}{l r r r r r}
        \toprule
\textbf{Bencher} &
\textbf{MAPE} & \textbf{Median} & \textbf{Q1} & \textbf{Q3} & \textbf{$\tau_{K}$}\\
\midrule
\llvmmca & 32.69\,\% & 23.43\,\% & 10.41\,\% & 51.62\,\% & 0.61 \\
\uica & 26.43\,\% & 15.59\,\% & 6.70\,\% & 42.39\,\% & 0.64  \\
\ithemal & 56.08\,\% & 46.20\,\% & 20.81\,\% & 74.30\,\% & 0.40 \\
\iaca & 27.48\,\% & 16.11\,\% & 6.70\,\% & 49.15\,\% & 0.64\\
\toolsensitivity{} & 20.27\,\% & 14.72\,\% & 5.84\,\% & 30.41\,\% & 0.82 \\
\bottomrule
    \end{tabular}}
    \caption{}
    \label{fig:overall_analysis_stats}
\end{subfigure}
\begin{subfigure}{\linewidth}
    \includegraphics[width=\linewidth]{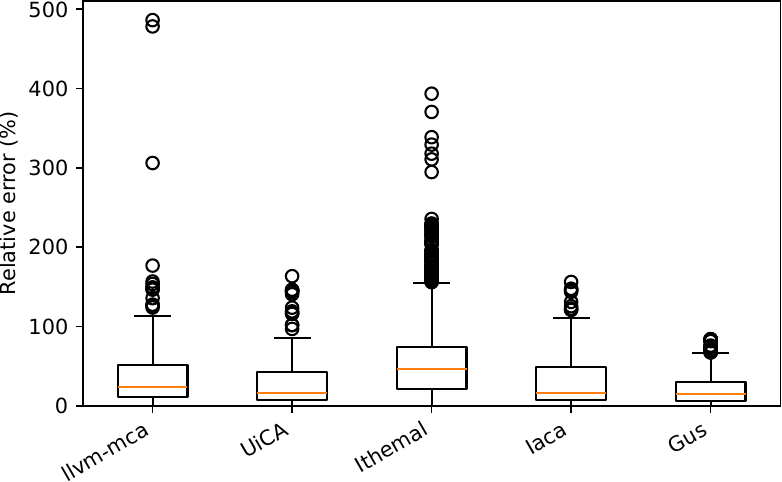}
    \caption{}
    \label{fig:overall_analysis_boxplot}
\end{subfigure}
	\caption{Statistical comparison (\subref{fig:overall_analysis_stats}) and corresponding error distribution (\subref{fig:overall_analysis_boxplot}) of throughput estimation tools on micro-benchmarks extracted from PolyBench~\cite{polybench}.}
	
\label{fig:overall_analysis_stats_and_boxplots}

\end{figure}

The error distributions of each code analyzer tools compared to ground truth measurements,
as well as synthetic statistical indicators are reported
in \figref{overall_analysis_stats_and_boxplots}. \toolsensitivity{} achieves the highest MAPE and Kendall
tau~\cite{kendall1938tau} ($\tau_{K} \in [0,1]$, which measures the fraction of
preserved pairwise ordering) scores. \uica{} and \iaca{} perform
similarly, faling just behind \toolsensitivity{}. \texttt{Ithemal} \cite{ithemal} is
the overall lesser accurate tool, with a MAPE of 56.08~\%.



\subsection{Bottleneck analysis}

We present in \autoref{tab:bottlenecks-polybench} a comparison of bottlenecks
measured on a dual socket Intel Xeon Gold 6230R CPU (Cascade Lake generation, Skylake-X microarchitecture)
by different methods over a set of micro-benchmarks extracted from our benchmarking harness.
The extracted subset
used a ``vanilla'' compilation strategy, using only the \texttt{-O3} and
\texttt{-march=skylake} flags, with no transformation from any of the loop nest
optimizers.

We compare several state-of-the-art tools for bottleneck analysis: code
analyzers (\uica~\cite{uica}, \iaca~\cite{iaca}), TAMd~\cite{topdown} and \texttt{Gus}.
For TAM, we used \toplev{}~\cite{toplev}, an
open-source implementation of TAM 4.7 for Intel
CPUs that builds upon \texttt{perf}~\cite{perf} to access hardware counters.

The \perf{}-based approach, contrary to code analyzers like \uica{} or \iaca{}
cannot limit the analysis to a basic block; and, contrary to
\toolsensitivity{}, cannot limit the analysis to a function call.
Thus, initialization/instrumentation wrappers used by the
PolyBench suite can mislead the former. To ensure a fair comparison,
we limited as much as possible this extra code.

As code analyzers typically work on a single basic block, reported bottlenecks
are the ones inferred for the hottest basic block of the kernel of each
benchmark. We assume that this is representative of the overall bottleneck of
the benchmark as our generation approach ensures that we tackle only the
loop whose execution time dominates the overall execution time.

Bottlenecks are given as is by the tools. If a tool does not provide a
bottleneck for a given benchmark, they are marked as \texttt{Unknown} in the
table. For \uica, we explain this by the arbitrary thresholds used to infer
bottlenecks (if the contribution to the total throughput of a given resource
doesn't exceed a given threshold, it is not reported as a bottleneck).

We also want to emphasize that a definite automatized groundtruth for
bottlenecks does not exist at the time of writing, as such our comparison
focuses on the strengths and weaknesses of each tool.

The retiring (\textit{Ret}) category is not referred as a bottleneck \textit{per se} by the TAM,
because it denotes when the pipeline is busy with typically useful operations.
Non-vectorized code might have a high retiring rate where it is considered as a
bottleneck, however we deal here with vectorized code where this rather
indicates that no bottlenecks have been found by the automatic analysis. We see
multiple such rows (\texttt{syr2k}, \texttt{correlation}, \texttt{covariance}) in the table where the TAM only
reports ``Retiring'' as a bottleneck and whereas code analyzers refine this to
a bottleneck. This category can otherwise correspond to a what's referred as
``Frontend'' (\textit{F-E}) bottleneck by code analyzers or \toolsensitivity.

Moreover, the TAM 4.7 does not make the distinction between what is referred as
dependencies-bound (\textit{Deps}) or latency-bound (\textit{Lat}) by code analyzers or \toolsensitivity{} and
the more general ``Backend'' (\textit{B-E}) category. This is the case for eleven rows, where
both \uica{} and \toolsensitivity{} agree on this ``Dependencies'' bottleneck
and the TAM rather reports ``Backend'' as a bottleneck. We are not aware of
any performance counters that would allow distinguishing between these two
categories, as such this shows the usefulness of code analyzers and
\toolsensitivity{} to provide more fine-grained bottlenecks in this case.

\toolsensitivity{} and \uica{} are able to provide multiple bottlenecks, this is
shown in the row of \texttt{jacobi-2d} where \toolsensitivity{} reports two
bottlenecks: latency and ROB ; or in the row of \texttt{gemm} where \uica{}
reports two bottlenecks: decoder and predecoder. This is an inherent limitation
for the underlying decision-tree used by \texttt{IACA 3.0}. The automated
implementation of the TAM 4.7 in \toplev{} only reports a single critical
bottleneck.
\begin{figure}
\begin{lstlisting}[language=C]
for(t6 = 1; t6 < 99; ++t6) {
    p[t4][t6] = -c/(a*p[t4][t6-1]+b);
    q[t4][t6] = (-d*u[t6][t4-1]+(1.0
        + 2.0*d)*u[t6][t4] - f*u[t6][t4+1]
        - a*q[t4][t6-1]) / (a*p[t4][t6-1]+b);
}
\end{lstlisting}
\caption{\texttt{adi} benchmark.}
\label{fig:adi}
\end{figure}

\begin{table}[t]
    \centering
    \resizebox{\columnwidth}{!}{%
    \begin{tabular}{lllll}
        \toprule
        Benchmark & \toplev & \uica & \iaca & \toolsensitivity{} \\
        \midrule
        2mm & B-E  (60.9\%)  & Deps  & Deps  & Lat (14.9\%)  \\
        3mm & B-E   (61.0\%) & Deps  & Deps  & Lat (14.9\%)  \\
        atax & B-E  (61.4\%) & Deps  & Deps  & Lat (15.0\%)  \\
        bicg & B-E  (61.3\%) & Deps  & Deps  & Lat (15.0\%)  \\
        doitgen & B-E  (34.2\%)  & Ports  & B-E   & \textcolor{gray}{Frontend (0.8\%)},  \textcolor{red!50}{\{P23, F-E, P4\}} \textcolor{red!50}{(14.9\%)}  \\
        mvt & B-E  (62.6\%) & Unknown  & F-E  & Lat (14.9\%)  \\
        gemver & Ret (77.3\%)  & Unknown  & B-E   & P23 (7.4\%)  \\
        gesummv & B-E  (60.5\%)  & Deps  & Deps  & Lat (15.0\%)  \\
        syr2k & Ret (74.0\%)  & Deps  & Deps  & Lat (15.0\%)  \\
        trmm & Ret (95.6\%) & Unknown & F-E  & F-E (14.9\%)  \\
        symm & B-E  (53.1\%)  & Deps  & B-E   & Lat (14.9\%)  \\
        syrk & B-E  (47.6\%) & Deps  & B-E   & Lat (14.9\%)  \\
        \hline
        gemm & Ret (78.4\%)  & \makecell[l]{Decoder \\ Predecoder}  & F-E  & F-E (14.9\%)  \\
        \hline
        gramschmidt & B-E  (55.0\%)  & Deps  & Deps  & Lat (14.9\%)  \\
        cholesky & B-E  (60.7\%)  & Deps  & Deps  & Lat (14.9\%)  \\
        \hline
        durbin & Ret (95.7\%)  & \makecell[l]{Decoder \\ Predecoder}  & F-E  & F-E (14.9\%)  \\
        \hline
        ludcmp & B-E  (74.4\%) & Deps  & B-E   & Lat (14.9\%)  \\
        trisolv & B-E  (60.8\%)  & Deps  & Deps  & Lat (15.0\%)  \\
        jacobi-1d & Ret (77.7\%)  & Issue, LSD  & B-E   & \textcolor{gray}{Frontend (4.8\%)}, \textcolor{red!50}{\{Lat,F-E\} (15.0\%)}  \\
        \hline
        heat-3d & Ret (79.3\%)  & Unknown  & F-E  & \makecell[l]{\textcolor{gray}{Lat (4.1\%), ROB (4.1\%), P4 (2.2\%)}  \\ \textcolor{red!50}{\{Lat, F-E, P01, P1, P23, P4\} (15.0\%)}}    \\
        \hline
        seidel-2d & B-E  (89.9\%)  & Deps  & B-E   & Lat (14.9\%)  \\
        fdtd-2d & B-E  (78.9\%)  & Ports  & F-E  & P23 (13.4\%)  \\
        jacobi-2d & B-E  (78.8\%)  & Unknown  & B-E & Lat (7.8\%), ROB (7.6\%)  \\
        adi & B-E  (79.5\%)  & Divider  & B-E   & Lat (14.9\%)  \\
        correlation & Ret (93.5\%)  & \makecell[l]{Decoder \\ Predecoder}  & F-E  & F-E (14.9\%)  \\
        covariance & Ret (78.8\%)  & Unknown  & F-E  & F-E (14.9\%)  \\
        floyd-warshall & F-E (43.6\%)  & LSD  & Deps  & F-E (14.9\%)  \\
        nussinov & Ret (97.3\%)  & LSD  & Deps  & F-E (14.9\%)  \\
        deriche & B-E  (64.7\%)  & Deps  & B-E   & Lat (14.9\%)  \\
        \bottomrule
        \end{tabular}}
    \caption{Comparison of bottleneck analysis tools on microbenchmarks extracted from PolyBench.
    Percentages in the \toolsensitivity{} column correspond to the upper-bound on the speedup
    with a 15\% resource acceleration (sub-1\% values omitted); percentages in the \toplev{} column
    represent the proportion of slots used by the resource.
    Gray bottlenecks for \toolsensitivity{} are found using sensitivity
    analysis on single resource, while red ones
    are found when varying multiple resources.}
    
    \label{tab:bottlenecks-polybench}
\end{table}

\toolsensitivity{} is able to provide the upper-bound on the speedup that can be
achieved by removing the bottleneck when accelerating the corresponding
resource. This allows to assess the potential of a given optimization. 






One apparent discrepancy between \toolsensitivity{} and \uica{} is shown in the
\texttt{adi} benchmark, whose C code is reported in \figref{adi}: \uica{} reports that the divider is the bottleneck,
whereas \toolsensitivity{} reports that the latency is the critical bottleneck.

A deeper analysis of the code shows that a dependency chain is present in the
loop body, indeed the previous \texttt{p} array value is used to compute the
next \texttt{p} array value, the same is true for the \texttt{q} array. Static
code analyzers such as \uica{} are oblivious to memory dependencies~\cite{anica}
in this basic block, as such they cannot detect the dominant latency bottleneck.
This highlights the usefulness of the dynamic nature of \toolsensitivity{} which
can easily detect this using a shadow memory.


Last but not least, we want to detail a category of benchmarks
(\texttt{heat-3d}, \texttt{dotigen}, \texttt{jacobi-1d}) that depict a certain
equilibrium between their resources. These benchmarks usually feature a very
high retiring rate (nearing 80\,\%) as reported by \toplev{} and are very well
optimized as the balance between their resources is very good. Optimizing a
single resource would not result in any significant speedup as this would not
affect the overall balance between the resources. This is shown by
\toolsensitivity{} single resource sensitivity analysis, where reported
bottlenecks feature a very low potential speedup (< 5\,\%). Optimizing for a
set of resources, can feature a higher potential speedup, we again demonstrate
this using \toolsensitivity{} multiple resource sensitivity analysis. Uncovering
this class of well-balanced benchmarks leads us to think that a more complete
framework for reasoning about bottlenecks is needed, especially for intertwined
resources.

\afterpage{\clearpage}


\section{Conclusion}

Identifying performance bottlenecks in programs that need to make the most of
the architecture is an increasingly critical task. We present
micro-architectural mechanisms on which program performance depends, as well as
the metrics and analysis methods that have been designed to guide optimization.
We propose to generalize bottleneck analysis by means of microarchitectural
sensitivity. It aims to automatically discover which resources influence a
program's overall performance by successively accelerating each of them and
observing the overall speedup generated by this variation. We present
\toolsensitivity{}, a dynamic code analyzer that implements sensitivity
analysis. We evaluate both its performance model and sensitivity analysis
algorithm on a set of microbenchmarks generated from the PolyBench benchmark
suite, and highlight its strengths and limitations compared with existing
methods based on hardware counters or performance models. \toolsensitivity{}'s
performance model achieves state-of-the-art accuracy in throughput estimation
over our experimental harness. As for sensitivity analysis, it allows us to
enhance the results of existing approaches, especially where
saturation is not sufficient to identify the one that limits parallelism.

\clearpage
\bibliographystyle{plain}
\bibliography{main}

\end{document}